\documentclass[journal]{IEEEtran}
\usepackage{cite}

\ifCLASSINFOpdf
   \usepackage[pdftex]{graphicx}
\else
   \usepackage[dvips]{graphicx}
\fi
\usepackage{amsmath}
\usepackage{algorithmic}

\usepackage{array}

\ifCLASSOPTIONcompsoc
  \usepackage[caption=false,font=normalsize,labelfont=sf,textfont=sf]{subfig}
\else
  \usepackage[caption=false,font=footnotesize]{subfig}
\fi
\usepackage{dblfloatfix}
\usepackage{url}

\usepackage{bbm}
\usepackage{amssymb}
\usepackage{amscd}
\usepackage[amsmath,thref,thmmarks]{ntheorem}
\theoremstyle{plain}

\newtheorem{theorem}{Theorem}
\newtheorem{definition}{Definition}
\newtheorem{lemma}{Lemma}

\usepackage{stmaryrd}

\usepackage{hyperref}

\begin{document}
%
\title{A Proof of Non-stationary Channel Polarization}
%
%
%

\author{Yizhi~Zhao, Hongmei Chi, Shiwei Xu, Lingjuan Wu and Yuling Fan 
\thanks{Y. Zhao was with the College of Informatics, Huazhong Agricultural University, Wuhan,
Hubei, China. E-mail: zhaoyz@mail.hzau.edu.cn.}
\thanks{H. Chi  was with the College of Science, Huazhong Agricultural University, Wuhan,
Hubei, China. E-mail: chihongmei@mail.hzau.edu.cn.}
}

\maketitle

\begin{abstract}
In this letter, we consider the proof of non-stationary channel polarization theory. First we construct a multi-channel stochastic process for the non-stationary channel polarization operation. Then based on this stochastic process, we extend Ar{\i}kan's standard martingale proof method on the average channel capacity and average channel Bhattacharyya parameter, by which we have proved the non-stationary channel polarization theory.
\end{abstract}


%
\IEEEpeerreviewmaketitle

\section{Introduction}

\IEEEPARstart{C}hannel polarization was first introduced by Ar{\i}kan in \cite{Arikan2009} that independent copies of channel $W$ can be polarized into noiseless channels and full-noise channels by the channel operation $\mathbf{G}_N$. Because of this polarization theory, polar codes have successfully solved the capacity-achieving problems of many stationary channel block (block formed with identical channels) models\cite{Arikan2009,Wyner1975,Mahdavifar2011,Vard2013strong,Wei2015,Gulcu2015}.

Then in \cite{Alsan2016}, the polarization of non-stationary channel block (block formed with different channels) was considered, and it was found that the non-stationary channel block can also be polarized by the channel operation $\mathbf{G}_N$. Later in \cite{Mahdavifar2018}, the achievablity of average symmetric capacity for non-stationary polarization was proved. This non-stationary polarization theory is a general extension of the original polarization theory which allows the application of polar codes on non-stationary models such as fading channel model\cite{Gopala2008}, arbitrarily varying WTC model\cite{Goldfeld2016}, etc.

In \cite{Arikan2009} Ar{\i}kan presented a standard proof for the channel polarization theory by constructing an stochastic tree-path process and using the concept the martingale. But in \cite{Alsan2016}, the non-stationary channel polarization theory was proved by a different method, which make us wonder whether Ar{\i}kan's method of martingale can also be used to prove the non-stationary polarization theory.

Therefore in this letter, we investigate the proof of non-stationary polarization theory by following Ar{\i}kan's standard method of martingale. We construct a multi-channel stochastic process for the non-stationary channel transformation. Based on this stochastic process, we consider the average channel capacity and average channel Bhattacharyya parameter at each level and extend the standard martingale method of \cite{Arikan2009}. Finally we proved that for non-stationary B-DMC sequence $W^{1:N}$ and generated channels $W_N^{(1:N)}$ by the non-stationary channel transformation $\mathbf{G}_N$, the fraction of indices $i\in[\![1,N]\!]$  for which $I(W_N^{(i)})\in (1-\delta,1]$ goes to $\mathbb{A}[I(W^{1:N })]$ and the fraction for which $I(W_N^{(i)})\in [0,\delta)$ goes to $1-\mathbb{A}[I(W^{1:N })]$ with any $\delta\in(0,1)$ and $N\rightarrow\infty$.

\section{Non-stationary Polarization}\label{sec_irg_polar}

\subsection{Notations}

We define the integer interval $[\![a,b]\!]$ as the integer set between $\lfloor a\rfloor$ and $\lceil b\rceil$. For $n\in \mathbb{N}$, define $N\triangleq 2^n$. Denote $X$, $Y$, $Z$,... random variables (RVs) taking values in alphabets $\mathcal{X}$, $\mathcal{Y}$, $\mathcal{Z}$,... and the sample values of these RVs are denoted by $x$, $y$, $z$,... respectively. Then $p_{XY}$ denotes the joint probability of $X$ and $Y$, and $p_X$, $p_Y$ denotes the marginal probabilities. Especially for channel $W$, the transition probability is defined as $W_{Y|X}$ and $W$ for simplicity. Also we denote a $N$ size vector $X^{1:N}\triangleq (X^1,X^2,...,X^N)$. When the context makes clear that we are dealing with vectors, we write $X^N$ in place of $X^{1:N}$. And for any index set $\mathcal{A}\subseteqq [\![1,N]\!]$, we define $X^{\mathcal{A}}\triangleq \{X^i\}_{i\in \mathcal{A}}$. For multi-block case, denote $X^{1:N}_{1:M}\triangleq (X^{1:N}_1,X^{1:N}_2,...,X^{1:N}_M)$. For the polar codes, we denote $\mathbf{G}_N$ the generator matrix , $\mathbf{R}$ the bit reverse matrix, $\mathbf{F}=
    \begin{bmatrix}\begin{smallmatrix}
        1 & 0 \\
        1 & 1
    \end{smallmatrix}\end{bmatrix}$
and $\otimes$ the Kronecker product, and we have $\mathbf{G}_N=\mathbf{R}\mathbf{F}^{\otimes n}$. Denote $\mathbb{A}[\cdot]$ as the average and $\mathbb{E}[\cdot]$ as the expectation.

\subsection{Polarization of Non-stationary Channels}

In Ar{\i}kan's channel polarization transformation, the whole combining and splitting operation can be broken recursively into single step $2\times2$ kernel transformation, as $(W,W)\mapsto(W^-,W^+)$, where $W^-:\mathcal{X}\rightarrow\tilde{\mathcal{Y}}$, $W^+:\mathcal{X}\rightarrow\mathcal{\tilde{Y}}\times\mathcal{X}$ and a one-to-one mapping $f:\mathcal{Y}^2\rightarrow\mathcal{\tilde{Y}}$.

Then for non-stationary channel cases, the framework of channel combining and splitting is the same as the polar channel transformation, known as $\mathbf{G}_N$. But the original initial channels $W^N$ is replaced by a block of non-stationary channels $W^{1:N}$ where $W^i:\mathcal{X}\rightarrow\mathcal{Y}$ with transition probability $W_{Y|X}^i$, and each $W^i$ is independent from others.

Thus we have the non-stationary $2\times2$ kernel transformation defined in Def.~\ref{def_ikt}

\begin{figure}[!h]
\centering
\subfloat[Polarized transformation]{\includegraphics[width=3cm]{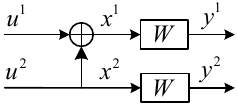}
\label{fig_rkt}}
\hfil
\subfloat[Non-stationary channel transformation]{\includegraphics[width=3cm]{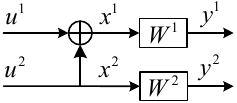}
\label{fig_ikt}}
\caption{Comparison of the $2\times2$ kernel between the polarized transformation and the non-stationary channel transformation.}
\label{fig_kt}
\end{figure}

\begin{definition}\label{def_ikt} Define the non-stationary $2\times2$ kernel transformation, illustrated in Fig.~\ref{fig_kt}, as $(W^1,W^2)\mapsto(W^-,W^+)$ which contains a channel operation pair $(\boxast,\circledast)$ that
\begin{equation}
W^-=W^1\boxast W^2\text{~and~}W^+=W^1\circledast W^2,
\end{equation}
specifically as
\begin{equation}
\begin{split}
&W^-(f(y^1,y^2)|u^1)=\sum_{u^2}\frac{1}{2}W^1(y^1|u^1\oplus u^2)W^2(y^2|u^2),\\
&W^+(f(y^1,y^2),u^1|u^2)=\frac{1}{2}W^1(y^1|u^1\oplus u^2)W^2(y^2|u^2).
\end{split}
\end{equation}
\end{definition}

By recursively performing the non-stationary $2\times2$ kernel transformation $(W^1,W^2)\mapsto(W^-,W^+)$ to the $N$ initial channels $W^{1:N}$, the structure of non-stationary channel transformation for generating the $W_N^{(1:N)}$ is defined as follows.

\begin{definition}\label{def_irs}(Non-stationary recursive structure) consider $N$ independent initial B-DMC $W^{1:N}$ with $N=2^n$, for all $q\in[\![0,n-1]\!]$, $Q=2^q$, $k\in[\![0,\frac{N}{2Q}-1]\!]$ and $i\in[\![2kQ+1,2kQ+Q]\!]$, by applying the non-stationary $2\times2$ kernel in Def.~\ref{def_ikt}, we have the recursive channel transformation as
\begin{equation}
(W_{Q}^{(i)},W_{Q}^{(i+Q)})\mapsto(W_{2Q}^{(2i-2kQ-1)},W_{2Q}^{(2i-2kQ)}),
\label{eq_recu}
\end{equation}
where
\begin{equation}
\begin{split}
&W_{2Q}^{(2i-2kQ-1)}=W_{Q}^{(i)}\boxast W_{Q}^{(i+Q)},\\
&W_{2Q}^{(2i-2kQ)}=W_{Q}^{(i)}\circledast W_{Q}^{(i+Q)},
\end{split}
\end{equation}
specifically as
\begin{equation}
\begin{split}
&W_{2Q}^{(2i-2kQ-1)}(y^{2kQ+1:2kQ+2Q},u^{2kQ+1:2i-2kQ-2}|u^{2i-2kQ-1})\\
&=\sum_{u^{2i-2kQ}}\frac{1}{2} W_{Q}^{(i)}(y^{2kQ+1:2kQ+Q},u_o^{2kQ+1:2i-2kQ-2}\\
&~~~\oplus u_e^{2kQ+1:2i-2kQ-2}|u^{2i-2kQ-1}\oplus u^{2i-2kQ})\\
&~~~\times W_{Q}^{(i+Q)}(y^{2kQ+Q+1:2kQ+2Q},u_e^{2kQ+1:2i-2kQ-2}|u^{2i-2kQ})
\end{split}
\end{equation}
and
\begin{equation}
\begin{split}
&W_{2Q}^{(2i-2kQ)}(y^{2kQ+1:2kQ+2Q},u^{2kQ+1:2i-2kQ-1}|u^{2i-2kQ})\\
&=\frac{1}{2} W_{Q}^{(i)}(y^{2kQ+1:2kQ+Q},u_o^{2kQ+1:2i-2kQ-2}\\
&~~~\oplus u_e^{2kQ+1:2i-2kQ-2}|u^{2i-2kQ-1}\oplus u^{2i-2kQ})\\
&~~~\times W_{Q}^{(i+Q)}(y^{2kQ+Q+1:2kQ+2Q},u_e^{2kQ+1:2i-2kQ-2}|u^{2i-2kQ}),
\end{split}
\end{equation}
where $u_o$ refers to odd term and $u_e$ refers to even term.
\end{definition}

\begin{figure}[!h]
\centering
\includegraphics[width=6cm]{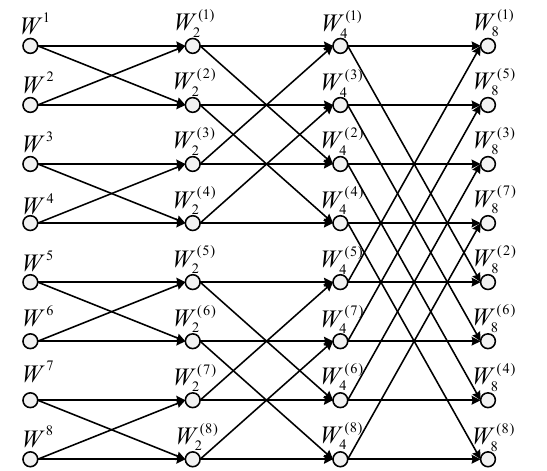}
\caption{The non-stationary channel transformation process from $W^{1:8}$ to $W^{(1:8)}_8$}
\label{fig_ctp}
\end{figure}

The recursive process of non-stationary transformation in Def.~\ref{def_irs} is illustrated in Fig.~\ref{fig_ctp} with $N=8$.

Then for the non-stationary transformation from $W^{1:N}$ to $W_N^{(1:N)}$, the generated channels are polarized as follows.

\begin{theorem}\label{theo_polarization} (Non-stationary polarization) For any non-stationary B-DMC block $W^{1:N}$, let $I(W)$ be the channel capacity of channel $W$ and $I(W^{1:N })$ be the sequence of $I(W^1),I(W^2),...,I(W^n)$. Assume that the expectation $\mathbb{E}[I(W^{1:N })]$ exists when $N\rightarrow\infty$, so that the average $\mathbb{A}[I(W^{1:N })]$ exists. Then the channels ${W_N^{(i)}}$ generated form $W^{1:N}$ by non-stationary channel transformation $\mathbf{G}_N$ are polarized as follows: for $N\rightarrow\infty$, $i\in[\![1,N]\!]$ and any fixed $\delta\in(0,1)$, have $I(W_N^{(i)})\in (1-\delta,1]$ with probability $\mathbb{A}[I(W^{1:N })]$ and $I(W_N^{(i)})\in [0,\delta)$ with probability $1-\mathbb{A}[I(W^{1:N })]$.
\end{theorem}

Note that in Theo.~\ref{theo_polarization}, we have made the assumption that $\mathbb{E}[I(W^{1:N })]$ exists when $N\rightarrow\infty$, which is not necessary in practical cases that the $N$ is finite.

\section{Proof of Non-stationary Polarization}\label{proof_polarization}

In this section, we present the proof of Theo.~\ref{theo_polarization} by following Ar{\i}kan's proof method of martingale in \cite[Section IV]{Arikan2009}. 

\begin{definition}\label{def_z}(\cite{Arikan2009}) For any given B-DMC $W:\mathcal{X}\rightarrow\mathcal{Y}$, the Bhattacharyya parameter is defined as
\begin{equation}
Z(W)\triangleq\sum_{y\in\mathcal{Y}}\sqrt{W(y|0)W(y|1)},
\end{equation}
which satisfies
\begin{equation}
\log\frac{2}{1+Z(W)}\leq I(W)\leq\sqrt{1-Z(W)^2},
\end{equation}
where $I(W)$ refers to the capacity of the channel $W$. Specially for BEC, have $I(W)=1-Z(W)$.
\end{definition}

\begin{lemma}\label{lem_i_irregular} For non-stationary $2\times2$ kernel transformation $(W^1,W^2)\mapsto(W^-,W^+)$, have
\begin{equation}
\begin{split}
&I(W^-)+I(W^+)=I(W^1)+I(W^2).\\
\end{split}
\end{equation}
\begin{IEEEproof} See Appendix~\ref{proof_i_irregular}.
\end{IEEEproof}
\end{lemma}

\begin{lemma}\label{lem_z_irregular} For non-stationary $2\times2$ kernel transformation $(W^1,W^2)\mapsto(W^-,W^+)$, have
\begin{equation}
\begin{split}
&Z(W^+)=Z(W^1)Z(W^2),\\
&Z(W^-)\leq Z(W^1)+Z(W^2)-Z(W^1)Z(W^2).\\
\end{split}
\end{equation}
And the second equal holds when $W^1$ and $W^2$ are BECs.
\begin{IEEEproof} See Appendix~\ref{proof_z_irregular}.
\end{IEEEproof}
\end{lemma}

As a corollary to Lem.~\ref{lem_z_irregular}, for BEC case, since $I(W)=1-Z(W)$, have
\begin{equation}
\begin{split}
&I(W^-)=I(W^1)I(W^2),\\
&I(W^+)= I(W^1)+I(W^2)-I(W^1)I(W^2).
\end{split}
\end{equation}

\begin{figure}[!h]
\centering
\includegraphics[width=8cm]{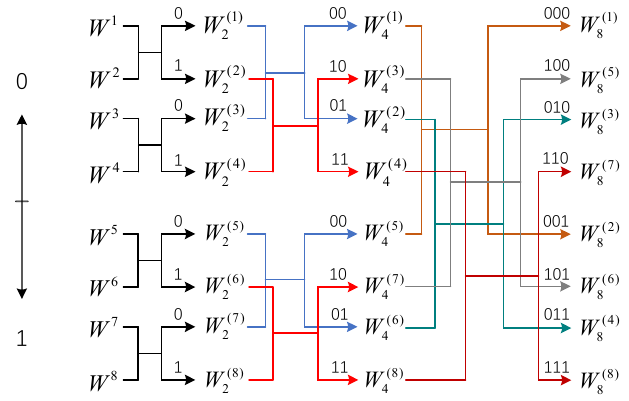}
\caption{The multi-channel stochastic process with $N=8$.}
\label{fig_mcsp}
\end{figure}

According to the recursive of non-stationary transformation in Fig.~\ref{fig_ctp}, we build \emph{a multi-channel stochastic process of the non-stationary channel transformation} which is illustrated in Fig.~\ref{fig_mcsp}.

Let $N=2^n$, $q\in[\![1,n]\!]$. If $q\rightarrow\infty$, we have $n\rightarrow\infty$. There are $n+1$ levels for the entire process form level $0$ to level $n$. Level $0$ is the initial level with $N$ channel variables $W^{1:N}$, also can be write as $W_0^{1:N}$. In the multi-channel stochastic process, all the initial channels are moving from level $0$ to level $n$ by \emph{a same random path} defined as $b^1b^2...b^n$. In the random path, $b^q$ is the moving direction from level $q-1$ to level $q$ randomly chosen from $\{0,1\}$ with probabilities $(\frac{1}{2},\frac{1}{2})$. $b^q=0$ refers the move to the \emph{upper node} at level $q$ and $b^q=1$ refers the move to the \emph{lower node} at level $q$. Define $W^i_q$ the channel variable at level $q$ corresponding to initial channel $W^i$. With the path information, $W^i_q$ can also be written as $W_{:b^1b^2...b^q}^{i}$. Define $W^{(1:N)}_{2^q}$ the $N$ possible channel values at level $q$.

Here is an example of the multi-channel stochastic process. Suppose $n=3$, the random path $b^1b^2b^3=001$. Then according to Fig.~\ref{fig_mcsp}, the moves of channels $W^{1:N}$ from level $1$ to level $3$ is:
\begin{equation}
\begin{split}
  \text{\emph{\textbf{Level 1:}}}&\\
  W_1^{1:8}=&W_{:0}^{1:8}\\
  =&(W_2^{(1)},W_2^{(1)},W_2^{(3)},W_2^{(3)},W_2^{(5)},W_2^{(5)},W_2^{(7)},W_2^{(7)});\\
  \text{\textbf{\emph{Level 2:}}}&\\
  W_2^{1:8}=&W_{:00}^{1:8}\\
  =&(W_4^{(1)},W_4^{(1)},W_4^{(1)},W_4^{(1)},W_4^{(5)},W_4^{(5)},W_4^{(5)},W_4^{(5)});\\
  \text{\textbf{\emph{Level 3:}}}&\\
  W_3^{1:8}=&W_{:001}^{1:8}\\
  =&(W_8^{(2)},W_8^{(2)},W_8^{(2)},W_8^{(2)},W_8^{(2)},W_8^{(2)},W_8^{(2)},W_8^{(2)}).
\end{split}
\end{equation}

Here is the reason for letting all the initial channels have a same random path in the multi-channel stochastic process. As described in Def.~\ref{def_irs} and illustrated in Fig.~\ref{fig_ctp}, for non-stationary channel transformation from $W^{1:N}$ to $W_N^{(1:N)}$, any generated channel $W_N^{(i)}$ is determined by all the initial channels $W^{1:N}$. Thus in the multi-channel stochastic process, for any channel $W_N^{(i)}$ at the last level, we should consider the effects of all the initial channels at level $0$, which means that we should consider the random case that all the initial channels end in a same node at the last level. And it is easy to observe from Fig.~\ref{fig_mcsp} that the only way for $W^{1:N}$ ending in a same node is to have a same random path.

For level $q$ with some random path $b^1b^2...b^q$, we define $\mathbb{A}[I_q^{1:N}]=\mathbb{A}[I(W^{1:N}_q)]$, $\mathbb{A}[Z_q^{1:N}]=\mathbb{A}[Z(W^{1:N}_q)]$. From Fig.~\ref{fig_mcsp}, we have that with the same random path $b^1b^2...b^n$, all the initial channels $W^{1:N}$ will move to a same node $W^{(\phi)}_{N}$ at level $n$, where $\phi$ could be any value within $[\![1,N]\!]$ which is calculated as $\phi=\sum_{q=1}^{n}b^q2^{n-q}+1$ according to different randomly chosen pathes. Thus we have $\mathbb{A}[I_n^{1:N}]= I_N^{(\phi)}$ and $\mathbb{A}[Z_n^{1:N}]=Z_N^{(\phi)}$ where $I_N^{(\phi)}$ represents $I(W^{(\phi)}_{N})$ and $Z_N^{(\phi)}$ represents $Z(W^{(\phi)}_{N})$. Then when $q,n\rightarrow\infty$, have $\mathbb{A}[I_\infty^{1:\infty}]=I_\infty^{(\phi)}$ and $\mathbb{A}[Z_\infty^{1:\infty}]=Z_\infty^{(\phi)}$.

Next, similar as \cite[Section IV]{Arikan2009}, define a probability space $(\Omega,\mathfrak{F},P)$. $\Omega$ is the space for $(b^1,b^2,...)\in\{0,1\}^\infty$. $\mathfrak{F}$ is the \emph{Borel field} generated by $S(b^1,...,b^q)\triangleq\{\omega\in\Omega:\omega^1=b^1,...,\omega^q=b^q\}$. $P$ is the probability measure defined on $\mathfrak{F}$ that $P(S(b^1,...,b^q))=1/2^q$. Then have $\mathfrak{F}^0\subset\mathfrak{F}^1\subset...\subset\mathfrak{F}^q$.

Let $K_q^{1:N}=W^{1:N}_{:b^1b^2...b^q}$ as the value of the multi-channel stochastic process that $K_q^i=W^i_{:b^1b^2...b^q}$. Denote i.i.d. RVs $\{B^q;q=1,2,...\}$ that $B^{1:q}$ takes on the sample value $b^1,...,b^q$. Then similar as \cite[Section IV]{Arikan2009}, the multi-channel stochastic process can be defined as follows. For $\omega=(\omega^1,\omega^2,...)\in\Omega$, $q\geq1$, define $B^q(\omega)=\omega^q$, $K_q^{1:N}(\omega)=W^{1:N}_{:\omega^1...\omega^q}$, $\mathbb{A}[I_q^{1:N}(\omega)]=\mathbb{A}[I^{1:N}(K_q^{1:N}(\omega))]$, $\mathbb{A}[Z_q^{1:N}(\omega)]=\mathbb{A}[Z^{1:N}(K_q^{1:N}(\omega))]$. And for $q=0$, let $K_0^{1:N}=W^{1:N}$, $\mathbb{A}[I_0^{1:N}]=\mathbb{A}[I(W^{1:N})]$, $\mathbb{A}[Z_0^{1:N}]=\mathbb{A}[Z(W^{1:N})]$. For any $q\geq0$, $B^q$, $K_q^{1:N}$, $\mathbb{A}[I_q^{1:N}]$ and $\mathbb{A}[Z_q^{1:N}]$ are measurable respect to BF $\mathfrak{F}$.

Consider the following expectation
\begin{equation}
\begin{split}
&\mathbb{E}\left[\mathbb{A}[I^{1:N}_{q+1}]|S(b^1,...b^q)\right]\\
&=\frac{1}{N}\sum_{i=1}^N\left[ \frac{1}{2}I(W^i_{:b^1...b^q0})+\frac{1}{2}I(W^i_{:b^1...b^q1})\right]\\
&\overset{(a)}{=}\frac{1}{N}\sum_{i=1}^N I(W^i_{:b^1...b^q})\\
&=\mathbb{A}[I^{1:N}_q],
\end{split}
\end{equation}
where $(a)$ is due to Lem.~\ref{lem_i_irregular} and all the $W^i$ will take the same path. Thus we have martingale $\{\mathbb{A}[I^{1:N}_q],\mathfrak{F}^q;q\geq0\}$ as
\begin{equation}
\begin{split}
&\mathfrak{F}^q\subset \mathfrak{F}^{q+1},~\mathbb{A}[I^{1:N}_q]~\text{is}~\mathfrak{F}^q\text{-measurable},\\
&\mathbb{E}\left[ |\mathbb{A}[I^{1:N}_q]| \right]<\infty,\\
&\mathbb{A}[I^{1:N}_q]=\mathbb{E}\left[ \mathbb{A}[I^{1:N}_{q+1}]|\mathfrak{F}^q \right].
\end{split}
\end{equation}
So have
\begin{equation}
\mathbb{E}[I_\infty^{(\phi)}]=\mathbb{E}\left[\mathbb{A}[I_\infty^{1:\infty}]\right]=\mathbb{A}[I_0^{1:\infty}].
\end{equation}

Similarly for expectation
\begin{equation}
\begin{split}
&\mathbb{E}\left[\mathbb{A}[Z^{1:N}_{q+1}]|S(b^1,...b^q)\right]\\
&=\frac{1}{N}\sum_{i=1}^N\left[ \frac{1}{2}Z(W^i_{:b^1...b^q0})+\frac{1}{2}Z(W^i_{:b^1...b^q1})\right]\\
&\overset{(a)}{\leq}\frac{1}{N}\sum_{i=1}^N Z(W^i_{:b^1...b^q})\\
&=\mathbb{A}[Z^{1:N}_q],
\end{split}
\label{eq_z_mart}
\end{equation}
where $(a)$ is due to Lem.~\ref{lem_z_irregular} and all the $W^i$ take the same path. Thus we have supermartingale $\{\mathbb{A}[Z^{1:N}_q],\mathfrak{F}^q;q\geq0\}$ as
\begin{equation}
\begin{split}
&\mathfrak{F}^q\subset \mathfrak{F}^{q+1},~\mathbb{A}[Z^{1:N}_q]~\text{is}~\mathfrak{F}^q\text{-measurable},\\
&\mathbb{E}\left[ |\mathbb{A}[Z^{1:N}_q]| \right]<\infty,\\
&\mathbb{A}[Z^{1:N}_q]\geq\mathbb{E}\left[ \mathbb{A}[Z^{1:N}_{q+1}]|\mathfrak{F}^q \right].
\end{split}
\end{equation}
So it converges a.e. and in $\mathcal{L}^1$ to a RV $\mathbb{A}[Z_\infty^{1:\infty}]$ such that
\begin{equation}
\mathbb{E}\left[|\mathbb{A}[Z_q^{1:N}]-\mathbb{A}[Z_\infty^{1:\infty}]|\right]\rightarrow\infty,
\end{equation}
followed with $\mathbb{E}\left[|\mathbb{A}[Z_{q+1}^{1:N}]-\mathbb{A}[Z_q^{1:N}]|\right]\rightarrow\infty$. From Lem.~\ref{lem_z_irregular}, have that for connected $i,j\in[\![1,N]\!]$, exists $i'\in[\![1,N]\!]$ that $Z^j_{q+1}=Z^{i}_qZ^{i'}_q$ with probability $1/2$ in the multi channel stochastic process. Besides, note that
\begin{equation}
\begin{split}
\mathbb{A}[Z_q^{1:N}Z_q^{1:N'}]&=\frac{1}{N}\sum_{i=1}^N Z_q^i Z_q^{i'}\\
&\overset{(a)}{\leq}\sqrt{\frac{1}{N}\sum_{i=1}^N(Z_q^i)^2\times\frac{1}{N}\sum_{i'=1}^N (Z_q^{i'})^2}\\
&\overset{(b)}{=}\mathbb{A}[(Z_q^{1:N})^2]
\end{split}
\label{eq_a_iql}
\end{equation}
where $(a)$ is due to Cauchy's inequality, and $(b)$ is due to $i$ are symmetric with $i'$ and $Z_q^i,Z_q^{i'}\in Z_q^{1:N}$. Thus have
\begin{equation}
\begin{split}
\mathbb{E}\left[|\mathbb{A}[Z_{q+1}^{1:N}]-\mathbb{A}[Z_q^{1:N}]|\right]&=\mathbb{E}\left[\frac{1}{N}|\sum_{i=1}^N Z_{q+1}^i- \sum_{i=1}^N Z_q^i|\right]\\
&=\mathbb{E}\left[\frac{1}{N}\sum_{i=1}^N |Z_{q+1}^i- Z_q^i|\right]\\
&\geq\frac{1}{2}\mathbb{E}\left[\frac{1}{N}\sum_{i=1}^N| Z_q^i Z_q^{i'}- Z_q^i|\right]\\
&=\frac{1}{2}\mathbb{E}\left[\mathbb{A}[Z_q^{1:N}]-\mathbb{A}[Z_q^{1:N}Z_q^{1:N'}]\right]\\
&\overset{(a)}{\geq}\frac{1}{2}\mathbb{E}\left[\mathbb{A}[Z_q^{1:N}]-\mathbb{A}[(Z_q^{1:N})^2]\right]\\
&=\frac{1}{2}\mathbb{E}\left[\mathbb{A}\left[Z_q^{1:N}(1-Z_q^{1:N})\right]\right]\geq0,
\end{split}
\end{equation}
where $(a)$ is due \eqref{eq_a_iql}. Since when $q\rightarrow\infty$, all the $Z_q^{1:N}$ will move to a same node $Z_\infty^{(\phi)}$, have $\mathbb{E}\left[Z_\infty^{(\phi)}(1-Z_\infty^{(\phi)})\right]=0$, which means that $Z_\infty^{(\phi)}$ equals $0$ or $1$. Form Def.~\ref{def_z}, have $I_\infty^{(\phi)}=1-Z_\infty^{(\phi)}$. Therefore, form $I_\infty^{(\phi)}=\mathbb{E}[I_\infty^{(\phi)}]=\mathbb{A}[I_0^{1:\infty}]$, easy to have
\begin{equation}
I_\infty^{(\phi)}=
\begin{cases}
1, ~\text{with probability} ~\mathbb{A}[I^{1:\infty }_0]\\
0, ~\text{with probability}~1-\mathbb{A}[I^{1:\infty }_0],
\end{cases}
\end{equation}
which completes the proof of Theo.~\ref{theo_polarization} for non-stationary polarization.

%

\section{Conclusion}\label{sec_con}

In this letter, we have presented a proof for the non-stationary polarization theory. We have constructed a multi-channel stochastic process for the non-stationary channel transformation. Based on this stochastic process, we study the average channel capacity and average channel Bhattacharyya parameter at each level and extend Ar{\i}kan's standard martingale proof method, by which we proved that for non-stationary B-DMC sequence $W^{1:N}$ and generated channels $W_N^{(1:N)}$ by the non-stationary channel transformation $\mathbf{G}_N$, the fraction of indices $i\in[\![1,N]\!]$  for which $I(W_N^{(i)})\in (1-\delta,1]$ goes to $\mathbb{A}[I(W^{1:N })]$ and the fraction for which $I(W_N^{(i)})\in [0,\delta)$ goes to $1-\mathbb{A}[I(W^{1:N })]$ with any $\delta\in(0,1)$ and $N\rightarrow\infty$.

\section*{Acknowledgment}

This work is supported in part by the National Natural Science Foundation of China (Grand No.62004077), the Natural Science Foundation of Hubei Province (Grant No.2019CFB137) and the Fundamental Research Funds for the Central Universities (Grant No.2662018JC007, 2662018QD057).

%
%
%

\ifCLASSOPTIONcaptionsoff
  \newpage
\fi




\begin{thebibliography}{1}

\bibitem{Arikan2009}
E. Ar{\i}kan, ``Channel polarization: a method for constructing capacity achieving codes for symmetric binary-input memoryless channels," \textit{IEEE Trans. Inf. Theory}, vol. 55, no. 7, pp. 3051-3073, Jul. 2009.

\bibitem{Wyner1975}
A. D. Wyner, ``The wire-tap channel," \textit{Bell System Tech. J.}, vol. 54, no. 8, pp. 1355-1387, Oct. 1975.

\bibitem{Mahdavifar2011}
H. Mahdavifar and A. Vardy, ``Achieving the secrecy capacity of Wiretap channels using Polar codes," \textit{IEEE Trans. Inf. Theory}, 2011, vol. 57, no. 10, pp. 6428-6443, Oct. 2011.

\bibitem{Vard2013strong}
E. \c{S}a\c{s}o\u{g}lu and A. Vardy, ``A new polar coding scheme for strong security on wiretap channels," \textit{IEEE Int. Symp. Inf. Theory (ISIT)}, pp. 1117-1121, Jul. 2013.

\bibitem{Wei2015}
Y.-P. Wei and S.Ulukus, ``Polar coding for the general wiretap channel," \textit{in Proc. Information Theory Workshop}, 1-5, Apr. 26/May 1 2015.

\bibitem{Gulcu2015}
T. C. Gulcu and A. Barg, ``Achieving secrecy capacity of the wiretap channel and broadcast channel with a confidential component," \textit{in Proc. IEEE Inf. Theory Workshop}, pp. 1-5, Apr. 26/May 1 2015.

\bibitem{Alsan2016}
M. Alsan and E. Telatar, ``A simple proof of polarization and polarization for non-stationary memoryless channels," \textit{IEEE Trans. Inf. Theory}, vol. 62, no. 9, pp. 4873-4878, Sep. 2016.

\bibitem{Mahdavifar2018}
Mahdavifar H., ``Fast polarization for non-stationary channels," \textit{IEEE Int. Symp. Inf. Theory (ISIT)}, pp. 849-853, June 2017. 


%


\bibitem{Gopala2008}
P. K. Gopala, L. Lai and H. El Gamal , ``On the secrecy capacity of fading channels," \textit{IEEE Trans. Inf. Theory}, vol. 54, no. 10, pp. 4687-4698, Oct. 2008.

\bibitem{Goldfeld2016}
Z. Goldfeld, P. Cuff and H. H. Permuter, ``Arbitrarily varying wiretap channels with type constrained states," \textit{IEEE Trans. Inf. Theory}, vol. 62, no. 12, pp. 7216-7244, Dec. 2016.


\end{thebibliography}
%

\section{Appendices}

\subsection{Proof of Lemma \ref{lem_i_irregular}}\label{proof_i_irregular}

Here we proof the Lem.\ref{lem_i_irregular}. Since $I(W^-)=I(U^1;Y^1Y^2)$, $I(W^+)=I(U^2;Y^1Y^2U^1)$, have
\begin{equation}
\begin{split}
I(W^-)+I(W^+)=&I(U^1;Y^1Y^2)+I(U^2;Y^1Y^2U^1)\\
\overset{(a)}{=}&I(U^1;Y^1Y^2)+I(U^2;Y^1Y^2|U^1)\\
=&I(U^1U^2;Y^1Y^2)\\
\overset{(b)}{=}&I(X^1X^2;Y^1Y^2)\\
\overset{(c)}{=}&I(X^1;Y^1)+I(X^2;Y^2)\\
=&I(W^1)+I(W^2),
\end{split}
\end{equation}
where $(a)$ and $(c)$ are because $U^1$ and $U^2$ are independent, $(b)$ is because the polar transform is a bijection.

\subsection{Proof of Lemma \ref{lem_z_irregular}}\label{proof_z_irregular}

Here we proof the Lem.\ref{lem_z_irregular}. For channel transformation of \eqref{eq_recu}, similarly as \cite[Appendix E]{Arikan2009}, we have
\begin{equation}
\begin{split}
&Z(W^+)\\
=&\sum_{y^1y^2u^1}\sqrt{W^+(f(y^1,y^2),u^1|0)W^+(f(y^1,y^2),u^1|0)}\\
=&\sum_{y^1y^2u^1}\frac{1}{2}\sqrt{W^1(y^1|u^1)W^2(y^2|0)}\sqrt{W^2(y^1|u^1\oplus1)W^2(y^2|1)}\\
=&\sum_{y^2}\sqrt{W^2(y^2|0)W^2(y^2|1)}\\
&\times\sum_{u^1}\frac{1}{2}\sum_{y^1}\sqrt{W^1(y^1|u^1)W^1(y^1|u^1\oplus1)}\\
=&Z(W^2)Z(W^1)
\end{split}
\end{equation}

\begin{equation}
\begin{split}
&Z(W^-)\\
=&\sum_{y^1y^2}\sqrt{W^-(f(y^1,y^2)|0)W^-(f(y^1,y^2|1))}\\
=&\sum_{y^1y^2}\frac{1}{2}\sqrt{W^1(y^1|0)W^2(y^2|0)+W^1(y^1|1)W^2(y^2|1)}\\
&\times\sqrt{W^1(y^1|0)W^2(y^2|1)+W^1(y^1|1)W^2(y^2|0)}\\
\overset{(a)}{\leq}&\sum_{y^1y^2}\frac{1}{2}\left[\sqrt{W^1(y^1|0)W^2(y^2|0)}+\sqrt{W^1(y^1|1)W^2(y^2|1)}\right]\\
&\times\left[\sqrt{W^1(y^1|0)W^2(y^2|1)}+\sqrt{W^1(y^1|1)W^2(y^2|0)}\right]\\
&-\sum_{y^1y^2}\sqrt{W^1(y^1|0)W^2(y^2|0)W^1(y^1|1)W^2(y^2|1)}\\
\overset{(b)}{=}&Z(W^1)+Z(W^2)-Z(W^1)Z(W^2)
\end{split}
\end{equation}
where $(a)$ is due to
\begin{equation}
\begin{split}
&\left[(\sqrt{ab}+\sqrt{cd})(\sqrt{ad}+\sqrt{cb})-2\sqrt{abcd}\right]^2\\
&=\left[\sqrt{(ab+cd)(ad+cb)}\right]^2+2(\sqrt{a}-\sqrt{c})^2(\sqrt{b}-\sqrt{d})^2\sqrt{abcd}
\end{split}
\end{equation}
Note that the equal of $(a)$ holds when $a=c$ or $b=d$ or $abcd=0$, which is satisfied when $W^1$ and $W^2$ are BECs. $(b)$ is due to
\begin{equation}
\begin{split}
\sum_{y^1y^2}&\frac{1}{2}\left[\sqrt{W^1(y^1|0)W^2(y^2|0)}+\sqrt{W^1(y^1|1)W^2(y^2|1)}\right]\\
&\times\left[\sqrt{W^1(y^1|0)W^2(y^2|1)}+\sqrt{W^1(y^1|1)W^2(y^2|0)}\right]\\
=&\frac{1}{2}\sum_{y^1}W^1(y^1|0)\sum_{y^2}\sqrt{W^2(y^2|0)W^2(y^2|1)}\\
&+\frac{1}{2}\sum_{y^1}W^1(y^1|1)\sum_{y^2}\sqrt{W^2(y^2|0)W^2(y^2|1)}\\
&+\frac{1}{2}\sum_{y^2}W^2(y^2|0)\sum_{y^1}\sqrt{W^1(y^1|0)W^2(y^1|1)}\\
&+\frac{1}{2}\sum_{y^2}W^2(y^2|1)\sum_{y^1}\sqrt{W^1(y^1|0)W^2(y^1|1)}\\
=&Z(W^1)+Z(W^2)
\end{split}
\end{equation}
and
\begin{equation}
\begin{split}
&\sum_{y^1y^2}\sqrt{W^1(y^1|0)W^2(y^2|0)W^1(y^1|1)W^2(y^2|1)}\\
&~~~~~~=\sum_{y^1}\sqrt{W^1(y^1|0)W^1(y^1|1)}\times\sum_{y^2}\sqrt{W^2(y^2|0)W^2(y^2|1)}\\
&~~~~~~=Z(W^1)Z(W^2)
\end{split}
\end{equation}

\end{document}